\documentclass[aps, prd, reprint, nofootinbib, longbibliography, preprintnumbers, amsfonts, amsmath, amssymb]{revtex4-1}
\pdfoutput=1 

\usepackage{ifpdf}
\usepackage[separate-uncertainty = true, multi-part-units = single,
            range-units = single]{siunitx}
\usepackage{physics}

\ifpdf
    \usepackage[pdftex, colorlinks = true, plainpages = false, pdfpagelabels]{hyperref}
    \usepackage[pdftex]{graphicx}
    \hypersetup{colorlinks = true}
\else
    \usepackage[colorlinks = true]{hyperref}
    \usepackage{graphicx}
\fi

\begin{document}


\title{The Excess Radio Background and Fast Radio Transients}

\author{John Kehayias}
\email{john.kehayias@vanderbilt.edu}
\affiliation{Department of Physics and Astronomy, Vanderbilt University\\
  Nashville, TN 37235, United States}

\author{Thomas W. Kephart}
\email{thomas.w.kephart@vanderbilt.edu}
\affiliation{Department of Physics and Astronomy, Vanderbilt University\\
  Nashville, TN 37235, United States}

\author{Thomas J. Weiler}
\email{tom.weiler@vanderbilt.edu}
\affiliation{Department of Physics and Astronomy, Vanderbilt University\\
  Nashville, TN 37235, United States}

\begin{abstract}

  \noindent
  In the last few years ARCADE~2, combined with older experiments, has
  detected an additional radio background, measured as a temperature
  and ranging in frequency from 22 MHz to 10 GHz, not accounted for by
  known radio sources and the cosmic microwave background. One type of
  source which has not been considered in the radio background is that
  of fast transients (those with event times much less than the
  observing time). We present a simple estimate, and a more detailed
  calculation, for the contribution of radio transients to the diffuse
  background. As a timely example, we estimate the contribution from
  the recently-discovered fast radio bursts (FRBs). Although their
  contribution is likely 6 or 7 orders of magnitude too small (though
  there are large uncertainties in FRB parameters) to account for the
  ARCADE~2 excess, our development is general and so can be applied to
  any fast transient sources, discovered or yet to be discovered. We
  estimate parameter values necessary for transient sources to
  noticeably contribute to the radio background.

\end{abstract}

\date{\today}

\maketitle

\section{Introduction}
\label{sec:intro}

Radio astronomy and astrophysics have an influential history,
intertwined with astronomy, cosmology, particle, and nuclear physics.
Today there are mysterious new sources being discovered in the radio
universe. In this work we will focus on two recent and exciting
developments, namely an observed and unexplained excess in the radio
background (ERB) and the discovery of fast radio bursts (FRBs). We use
the latter fast radio transient (FRT) sources as the example for our
modeling.

The ARCADE~2 (Absolute Radiometer for Cosmology, Astrophysics and
Diffuse Emission) experiment reported~\cite{arcade2} an excess in
temperature, after subtracting the cosmic microwave background (CMB)
and accounting for other known backgrounds, in measurements at
frequencies from \SIrange[range-phrase = --]{\sim 3}{10}{\GHz}. When
combined with previous data and analysis in the literature there is an
excess that extends from \SI{22}{MHz} to nearly \SI{10}{GHz}. This is
well fit by a power law in absolute temperature versus frequency with
the following form~\cite{arcade2}
\begin{equation}
  \label{eq:exradio}
  T = \left(\SI{24.1 \pm 2.1}{K}\right) \left(\frac{\nu}{\SI{310}{MHz}}\right)^{\num{-2.599 \pm 0.0366}}.
\end{equation}

\begin{figure}
  \centering
  \includegraphics[width=\columnwidth]{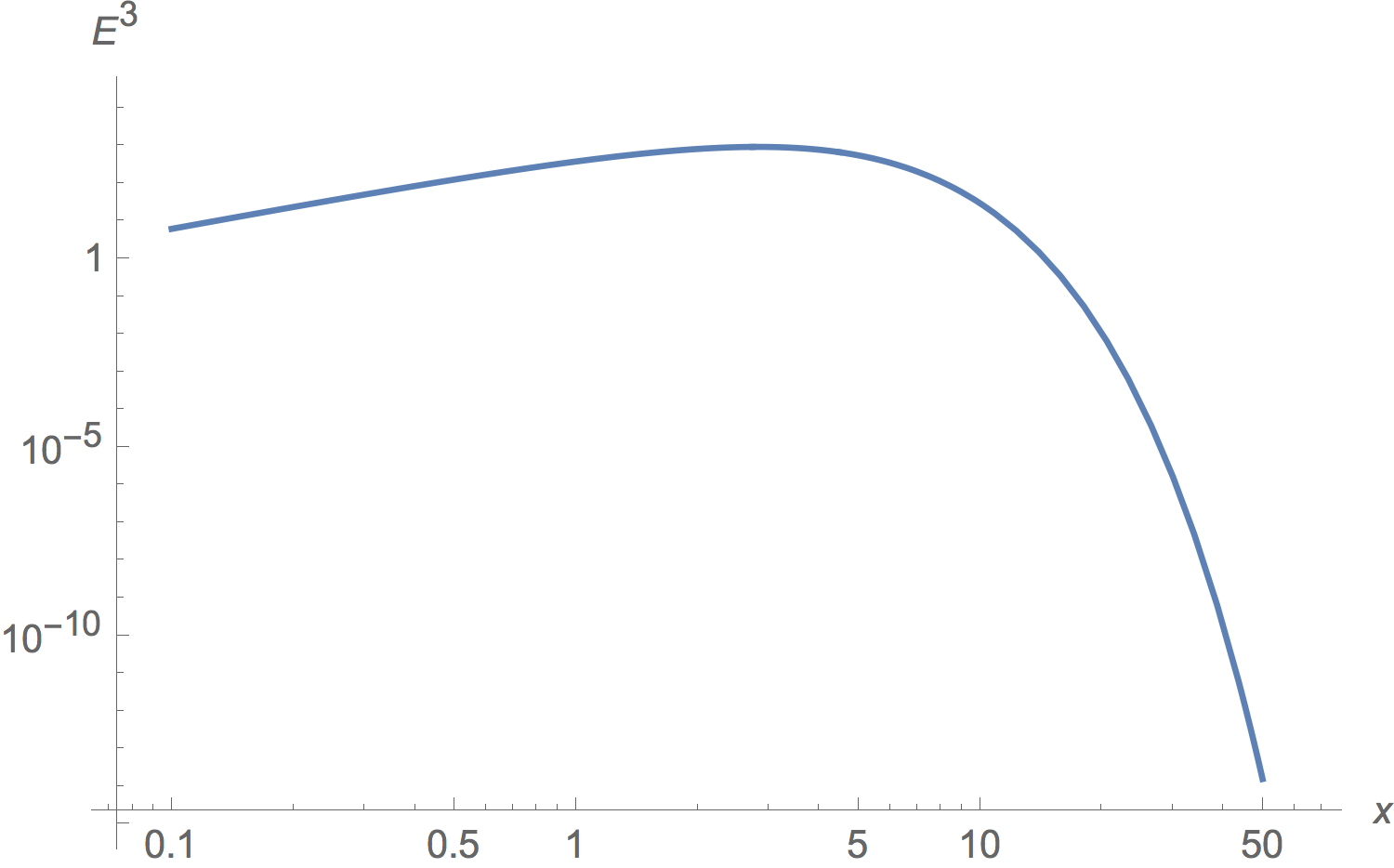}
  \caption{A log-log plot of eq.~\eqref{eq:bosedistrb} versus $x$,
    with $T = 10$.}
  \label{fig:bose}
\end{figure}

While ARCADE~2 measures a temperature for the radio background,
transient sources like FRBs are measured by other experiments in units
of spectral irradiance or spectral flux density: Janskys,\footnote{For
  our fellow particle physicists, one might be tempted to identify the
  energy implicit in Watts with that in the Hertz in the denominator,
  but these energies do not ``cancel.'' One should think of the Watts
  as the power that is received by the radio telescope at a particular
  frequency in Hertz (and per unit area). The bandwidth (frequency
  range of observation), measured in Hertz as well as the observing
  area, will directly affect the power received, and thus it is more
  sensible to express the radio unit as power per unit frequency and
  area (an energy flux density): the Jansky.} where
$\SI{1}{Jy} \equiv \SI{e-26}{\watt\per\meter\squared\per\hertz}$. To
convert these temperature measurements into a flux density, we show
the standard distribution function (i.e.~Planck radiation law) for a
massless Bose particle (here a photon) as a function of the
dimensionless variable $x \equiv h\nu/k_BT$ in Fig.~\ref{fig:bose}.
The energy density per frequency distribution, $\dv*{\rho}{\nu}$, is
given by
\begin{equation}
  \label{eq:bosedistrb}
  \dv{\rho}{\nu} = \frac{2}{\pi} \frac{T^3 x^3}{e^x - 1} = \frac{2h\nu^3}{c^2}\frac{1}{e^\frac{h\nu}{k_BT} - 1},
\end{equation}
with the first form written in natural units ($\hbar = c = k_B = 1$)
with overall units of \SI{}{energy^3}, and the second is the typical
textbook form with the constants restored and units proportional to
\SI{}{Janskys}. Note that when $T$ is allowed to vary with frequency,
as in the ERB data, then the spectral form will differ from the fixed
distribution shown in Figure~\ref{fig:bose}.

This excess background has persisted and has been difficult to
understand despite different analyses (see,
e.g.,~\cite{Singal11122010, seiffert}), and remains of unknown origin
(for a recent study see, e.g.,~\cite{radiobkgrevist} and references
therein). If the background comes from additional extragalactic
sources, they must be very numerous and faint,
e.g.~\cite{Singal11122010, Vernstrom21082011}. Additionally, the
background appears to be extremely smooth, which puts further
constraints on the size and cosmological distribution of such
background sources~\cite{holder}. There have been several different
proposals for an alternative origin for this radio background,
including dark matter~\cite{arcadebkgmodel1, *arcadebkgmodel2,
  *arcadebkgmodel3, *arcadebkgmodel4, *arcadebkgmodel5,
  *arcadebkgmodel6}, but the question is still open.

More recently there has been the discovery, first with the Parkes
telescope~\cite{parkes1, parkes2, parkes3} and then at
Arecibo~\cite{arecibo}, of a new class of radio objects, dubbed ``fast
radio bursts'' (FRBs) due to their only known activity occurring on
the millisecond Earth-time scale (FRB discoveries have continued as
well~\cite{thorntonthesis, *Burke-Spolaor:2014rqa, *Petroff:2014taa,
  *Ravi:2014mma}). Based upon their dispersion measures
(DMs),\footnote{Without candidate source galaxies for FRBs thus far,
  the only way to estimate their distance is through their dispersion
  measure. The DM is determined by the electron density along the line
  of sight, which delays lower frequency signals compared to higher
  frequencies. Given an estimate for the electron contribution from
  the galaxy, the large DMs of FRBs lead to the inference of
  extragalactic sources.} these FRBs are believed to be extragalactic
in origin (although this has been challenged, see, e.g.,
\cite{Kulkarni:2014vea, *loebgalactic}). The physical process powering
an FRB is so far unknown, with proposals ranging from neturon
stars~\cite{neutron1, *neutron2}, axions~\cite{axionfrb1, *axionfrb2},
and black holes~\cite{parkes2, Lipunov:2013axa} to several
others~\cite{Popov:2013bia, *Kashiyama:2013gza, *Connor:2015era,
  *Katz:2015mpa}. It is further inferred that these unknown objects
release \SIrange[range-phrase = --]{\sim e38}{e40}{erg} during their
millisecond duration, corresponding to a typical peak flux density of
\SI{> 0.4}{Jy}. While there are only a few observed FRBs (\num{11} at
this moment; an up to date list with measured parameters and
references is available at \url{http://astro.phys.wvu.edu/FRBs/}), a
rough estimate of the rate for these events gives of order
\num[group-separator = {,}]{10000} per day in the whole sky out to
$z = 0.75$ (recent studies, such as~\cite{Rane:2015sxa}, put the rate
a bit lower, though still consistent with earlier results and of
negligible effect for our present analysis).

Motivated by FRBs, we will consider general fast transient radio
sources as contributing to the ERB\@. By ``fast'' we mean that the
timescale of such an event is much less than the time of observations
(which is of order an hour for ARCADE~2). Such sources have not been
considered in the past for calculating the diffuse radio background;
we will adjust the steady state radio source calculation to
accommodate the transient contribution.

This paper is organized as follows. First, in
Section~\ref{sec:estimate}, we will introduce the basic calculation
for radio source counts for the radio background, and modify it to
account for fast transients. While the FRB contribution to the excess
radio background is several orders of magnitude too small to make
detectable contribution, a more detailed calculation for transient
sources, e.g.~similar to the calculation for supernova neutrinos, may
be useful. In Section~\ref{sec:detailedcalc} we present a more
complete analysis. We then compare a general fast source to the excess
background to derive approximate parameters for such a source to
contribute to the excess at the percent level or more in
Section~\ref{sec:fitting}. Finally, in Sections~\ref{sec:discussion}
and \ref{sec:conclusions} we discuss our results and future directions
before concluding.

\section{A Radio Background from Fast Transients}
\label{sec:bkg}

Fast transient sources have not been considered as a contribution to
the observed ERB\@. One might ask if they can contribute appreciably
to the radio background. In this Section we will first use a
simplified analysis to estimate the contribution from fast transients
before providing a more comprehensive calculation. Fast transient
sources need to be treated slightly differently in the usual radio
background calculation in order to properly account for their event
rate and individual event time. For a recent review of FRT searches
and related bounds, see~\cite{Fender:2015sca} and references therein.

\subsection{An Estimate}
\label{sec:estimate}
For a typical source (static, or slowly changing, compared to typical
observation times), the contribution to the cosmic radio background
(CRB) (see e.g.~\cite{Singal11122010}) is
\begin{equation}
  \label{eq:bkgdnds}
  B_\text{CRB} = \int_{S_\text{min}}^\infty S \dv{N}{S} \dd{S},
\end{equation}
where $S$ is the flux density at Earth
(energy/(time$\cdot$area$\cdot$frequency), e.g.~in units of Janskys),
and $\dv{N}{S}$ is the differential number of sources at that flux
density. The unit for radio observations, the Jansky, implicitly has a
time such that multiplying by an observation time, area, and frequency
bandwidth, yields the total energy observed.

For transient sources, however, there are now two timescales which are
introduced: the duration of the source emission and the inverse rate
of sources appearing in the sky. For fast sources (where ``fast''
means short compared to the observation time, e.g.~ARCADE~2 here), the
entire energy of each event is observed. Instead of a flux density per
source, we should use an energy density: we multiply $S$ by $\var{t}$,
the time of the source mission. We still want the same units at the
end, so we need an inverse time, which comes from the number per flux
density being replaced by a rate per flux density (i.e.~a number per
energy density), $\dv*{N}{S}\dd{t}$, with units of per second after
integration over $S$. In other words, the source flux density is to be
multiplied by the duration of the transient source, while the number
density becomes a rate. We can write this as
\begin{equation}
  \label{eq:bkgdndnstrans}
  B_\text{CRB}^\text{FRT} = \int_{S_\text{min}}^{\infty} S\var{t} \frac{\dd[2]{N}}{\dd{S}\dd{t}} \dd{S},
\end{equation}
The result can properly be interpreted as the incoming energy flux
density, just as in Eq.~\eqref{eq:bkgdnds}.\footnote{Implicitly this
  is over the entire sky, but one can divide by $4\pi$ or else use a
  rate which is per steradian.}

Let us consider observations at one particular frequency, \SI{\sim
  1.4}{\GHz}, for which there is observation of both the
ERB\footnote{For now we will ignore any errors in the measurements,
  though the ERB data at \SI{1.4}{\GHz} is consistent (within errors)
  with zero after additional source subtractions. One could use a
  different frequency for the ERB, with an assumption of the FRB
  spectrum, and would find similar results.} and FRBs. The radio
background can be expressed as \SI{e4}{Jy/sr} (e.g.,
see~\cite{Singal11122010}). The estimated event rate of FRBs of $10^4$
per day in the whole sky (for $S > \SI{1}{Jy}$) corresponds to
$\SI{e-2}{\per\second\per\steradian}$ in units more useful for us. As
a lower estimate on the FRB contribution to the ERB at \SI{1.4}{\GHz},
we simply multiply this rate by the typical flux density times the
millisecond duration, $\SI{1}{Jy} \times \SI{e-3}{\second}$. Thus the
FRB contribution has a lower bound of $\sim 10^{-9}$ times the
observed radio background; this is a lower bound because the true
contribution will be larger, from inclusion of further/dimmer events,
as well as modeling the distribution of FRBs in redshift. (We will see
in Section~\ref{sec:frbmodel} that our more detailed calculation,
using a model for the FRB spectrum and distribution, increases the FRB
background by a factor of \num{\sim 100}.)

To verify this very rough estimate, we will assume a form for the
differential number per flux density, and integrate backwards in time
(outwards in space). (Since it is estimated from the DMs that the
observed FRBs have red-shifts $\lesssim 1$, the red-shifting of
energies, and dilation of time, are small.) With the assumptions that
FRBs are standard candles observed at extragalactic distances (as is
so far consistent with all available data), the nearby ($z < 1$ or
$S > \SI{1}{Jy}$) distribution approximately follows that of a static
Euclidean universe, $\dv{N}{S} = S^{-5/2}$. We normalize this
distribution to the estimated FRB rate above,
$\int_1^{\infty}\dv{N}{S}\dd{S} =
\SI{e-2}{\per\second\per\steradian}$.
Inputting this assumption allows integration of
eq.~\eqref{eq:bkgdnds}, and we find the FRB contribution matches our
rough estimate of \num{e-9} of the observed radio excess.

\subsection{A More Detailed Calculation}
\label{sec:detailedcalc}
To calculate the radio background due to FRTs in the universe, we
follow the same basic procedure as in, e.g., the neutrino background
from supernovas (see~\cite{Lunardini:2010ab} for a review). The two
main ingredients are an event rate (number per volume per time)
evolving in redshift, $R(z)$, and the spectrum (energy per frequency)
of an individual event, $F(\nu)$, with $\nu$ the frequency observed at
Earth. The incoming isotropic flux background, in units of
energy/(time $\cdot$ area $\cdot$ steradian $\cdot$ frequency), or
\SI{}{Jy\per\steradian}, is then calculated by the following integral,
\begin{equation}
  \label{eq:bkg}
  B(\nu) = \frac{1}{4\pi}\frac{c}{H_0}\int_0^{z_\text{max}}\mathrm{d}z \frac{R(z)F(\nu')}{\sqrt{\Omega_\text{m}(1 + z)^3 + \Omega_\Lambda}},
\end{equation}
where $c/H_0$ is the Hubble length, $z_\text{max}$ is the redshift
back to which we are integrating sources, $\nu' \equiv \nu(1 + z)$ is
the blueshifted frequency, and $\Omega_\text{m}$ and $\Omega_\Lambda$
are the energy densities of matter and dark energy, respectively.

\begin{figure}
  \centering
  \includegraphics[width=\columnwidth]{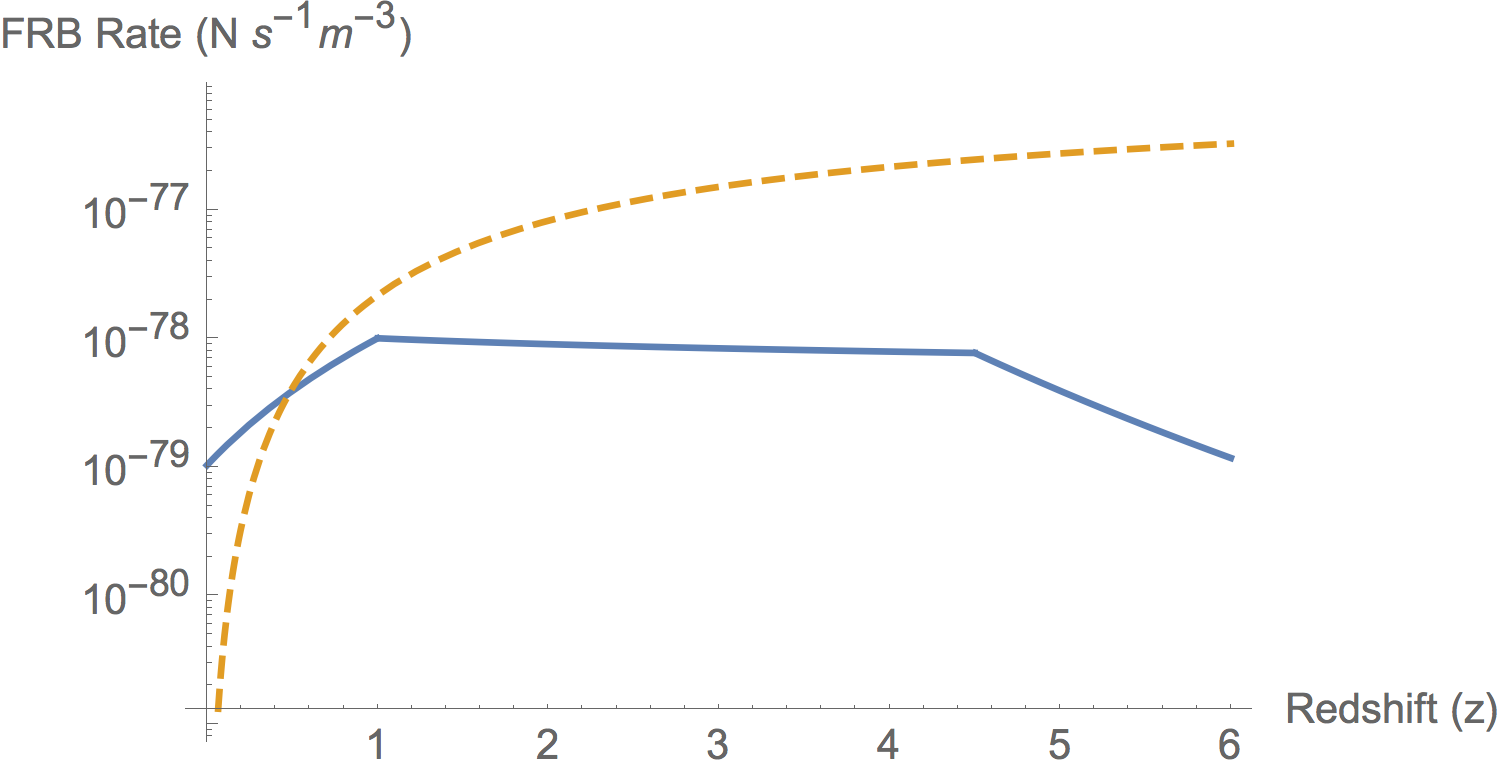}
  \caption{A model for the FRB rate scaled with comoving volume
    (orange, dashed) and tracking the star formation rate (blue,
    solid). The rates are normalized such that their integral to
    $z = 0.75$ matches the approximate observed rate.}
  \label{fig:frbratecomv}
\end{figure}

To perform this calculation we will need to model our radio sources.
First, let us consider how the sources are distributed in redshift. We
will employ two models, one where the source distribution simply
scales with the comoving volume as a standard candle (FRBs were
modeled like this in~\cite{Lorimer21112013}), and another which tracks
the star formation rate. These models are shown in
Figure~\ref{fig:frbratecomv}.

In the first model we assume the sources have a constant number
density in comoving volume, then the rate simply scales with the
fraction of comoving volume out to $t_f(z)$. With
\begin{equation}
  \label{eq:covol}
  D(z) \equiv c\int_{t_0}^{t_f}\frac{\dd{t}}{a(t)} = \frac{c}{H_0}\int_0^z \mathrm{d}z' \frac{1}{\sqrt{\Omega_\mathrm{m}(1 + z')^3 + \Omega_\Lambda}}
\end{equation}
the comoving distance, the source rate is
\begin{equation}
  \label{eq:covolrate2}
  R_\text{CoV}^\text{FRT}(z) = N_\text{CoV}^\text{FRT}\frac{D(z)^3}{D(z_*)^3},
\end{equation}
where $N_\text{CoV}^\text{FRT} = R_\text{CoV}^\text{FRT}(z_*)$ is a
normalization to be fixed by the source model or observation.

Alternatively, we model the source rate to be proportional to the star
formation rate. There are several different parameterizations and we
follow the one formulated in Table II of~\cite{Hopkins:2006bw}, but
with some parameters slightly rounded as in~\cite{Lunardini:2010ab},
\begin{equation}
  \label{eq:sfrate}
  R_\text{star}(z) \propto
    \begin{cases}
      (1 + z)^{3.28},      &\quad z < 1\\
      a_1 (1 + z)^{-0.26}, &\quad 1 < z < 4.5\\
      a_2 (1 + z)^{-7.8},  &\quad 4.5 < z
    \end{cases},
\end{equation}
with $a_1 = 11.6$ and $a_2 = \num{4.45e6}$ fixed by continuity. The
transient radio source rate is then
\begin{equation}
  \label{eq:sffrbrate}
  R_\text{SF}^\text{FRT}(z) = N_\text{SF}^\text{FRT} R_\text{star}(z),
\end{equation}
with a normalization denoted by $N_\text{SF}^\text{FRT}$.

Now we turn to modeling the spectral properties of the fast sources
themselves. As the simplest model we will take the spectrum of the
sources to follow a power law (in frequency) over some range of
frequency. The free parameters will be the spectral index, overall
normalization, and frequency range. Since we are studying the radio
background over observations much longer than the source timescale, we
will not need to directly assume this timescale. Instead, it is
implicit in our normalization of the total energy. For simplicity we
will assume the sources are either standard candles or consider all
quantities to be typical or average values.

We write the spectrum as
\begin{equation}
  \label{eq:specpowerlaw}
  F(\nu) = F_\text{norm} \left( \frac{\nu}{\nu_*}\right)^\alpha,
\end{equation}
with spectral index $\alpha$, reference frequency $\nu_*$, and
normalization $E_\text{norm}$ with units of energy per unit frequency.
The normalization is fixed by integrating the spectrum over its
frequency range, $\nu_\text{min}$ to $\nu_\text{max}$, to determine
the total energy released in an event,
\begin{equation}
  \label{eq:specnorm}
  E_\text{total} = \int_{\nu_\text{min}}^{\nu_\text{max}}\dd{\nu}F(\nu).
\end{equation}

\subsubsection{Fast Radio Burst Modeling}
\label{sec:frbmodel}
Before considering a general (as yet unknown) source to fit to the
excess radio background, we return to FRBs as an example of applying
the above basic modeling. With the current data on FRBs, the consensus
on the approximate rate of FRBs is about \num[group-separator =
{,}]{10000} per day in the whole sky, out to a redshift \num{0.75}. We
will use this observed rate, expressed in more useful units as
\begin{equation}
  \label{eq:frb075}
  R_\text{FRB}(z\le 0.75) \equiv \SI{2.40e-79}{\per\second\per\cubic\meter},
\end{equation}
as the integrated rate to redshift \num{0.75} to fix the
normalization.

If we assume FRBs are standard candles with a constant number density
in comoving volume (as in~\cite{Lorimer21112013}), their rate is
modeled by eq.~\eqref{eq:covolrate2} with the normalization,
$N_\text{CoV}^\text{FRB} \equiv 4.68 R_\text{FRB}(z \le 0.75)$
determined by setting $z_* = 0.75$ in eq.~\eqref{eq:covolrate2} and
integrating out to $z_*$. In our alternative model based on the star
formation rate, eq.~\eqref{eq:sfrate}, the normalization, obtained by
direct integration, is
$N_\text{SF}^\text{FRB} \equiv 0.43 R_\text{FRB}(z \le 0.75)$.

The second set of parameters for FRBs is their \textit{typical} energy
output and spectrum. From the observed events thus far, the energy
released is about \SIrange{e31}{e33}{\J}, or \SIrange{e38}{e40}{erg}.
We use the FRB model in~\cite{Lorimer21112013} as a reference point: a
spectral index of \num{-1.4} from \SI{10}{\MHz} to \SI{10}{\GHz}, with
a bolometric luminosity of \SI{8e44}{erg\per\s}, or a total energy of
\SI{8e34}{\J} (using an event time of \SI{1}{\milli\second}).

With these parameters we can calculate the FRB contribution to the
radio background, again at \SI{1.4}{\GHz}. We find that the FRB
contribution is approximately \num{3e-7} of the observed excess with a
constant comoving volume event distribution, and \num{0.5e-7} when
using an event rate proportional to the star formation rate.

Given the large uncertainties in what is known about FRBs, we can vary
the parameters to some degree. From eq.~\eqref{eq:bkg} we know that
any changes to the overall rate or energy of the radio transients will
proportionally change the calculated background contribution. For
instance, if the rate is underestimated by a factor of a few and the
total energy is underestimated by an order of magnitude (both
reasonable possibilities, given the observed values and error bars),
the FRB background contribution could be \num{e-5} of the excess.
While further tuning of parameters or modeling might improve this
slightly, it seems unlikely that, without a drastic modification of
what is known about FRBs, could they make up a significant proportion
of the ERB\@.

\section{Fitting a Fast Transient Source to the Excess Background}
\label{sec:fitting}

We will now fit the flux density of a new source of FRTs to the
observed ERB from ARCADE~2 and previous experiments. These experiments
report a temperature beyond the contribution from the CMB temperature
of \SI{2.725}{K} and the galactic foreground. We use ``Fit 1''
of~\cite{gervasi} to estimate the contribution from unresolved
extragalactic discrete sources (see also~\cite{condon} for resolving
such sources). Using the Planck radiation law,
Eq.~\eqref{eq:bosedistrb}, we convert these temperatures to an
expected spectral flux density, in units of
\SI{}{Jy/sr}.\footnote{Note that the reported errors in temperature
  become asymmetric in radiance; we use the smaller of the max/min
  errors for each data point as a common error for fitting purposes.}
The resulting data is shown in Table~\ref{tab:data}. We note that the
data point at \SI{1.4}{\GHz} is consistent with zero given the errors
on the measured temperature after the above subtractions. We do not
include data at frequencies larger than \SI{\sim 8}{\GHz} as they are
also consistent with zero.

\begin{table*}[htbp]
  \begin{tabular}{|c|c|c|c|}
    \hline
    Frequency & Temperature (\SI{}{K}) & Final Temperature (\SI{}{K}) & Flux Density (\SI{}{\kilo Jy/sr})\\
    \hline
    \SI{22}  {\MHz} & \num{20355 \pm 5181}  & \num{13479}   & \num{200.430 \pm 77.043}\\
    \SI{45}  {\MHz} & \num{3864 \pm 502}    & \num{2866}    & \num{178.300 \pm 31.2}  \\
    \SI{408} {\MHz} & \num{13.42 \pm 3.52}  & \num{8.107}   & \num{41.410 \pm 18.0}   \\
    \SI{1.42}{\GHz} & \num{3.271 \pm 0.526} & \num{0.4568}  & \num{26.24 \pm 32.6}    \\
    \SI{3.20}{\GHz} & \num{2.787 \pm 0.010} & \num{0.05205} & \num{2.667 \pm 1.38}    \\
    \SI{3.41}{\GHz} & \num{2.770 \pm 0.008} & \num{0.03662} & \num{0.6777 \pm 0.485}  \\
    \SI{7.98}{\GHz} & \num{2.761 \pm 0.013} & \num{0.03516} & \num{13.92 \pm 13.9 e-3}    \\
    \SI{8.33}{\GHz} & \num{2.743 \pm 0.015} & \num{0.01725} & \num{73.23 \pm 73.2 e-6}\\
    \hline
  \end{tabular}
  \caption{\label{tab:data}The ERB data used in this analysis. The data is from ARCADE~2 and
    older experiments (see Table~4 of Ref.~\cite{arcade2} and references
    therein), with the first temperature being the reported temperature
    before subtractions, but accounting for the galactic foreground. The
    final temperature is after the CMB, and extragalactic sources have
    been subtracted (the error bars are the same as in the previous column).
    The flux density is derived from this final temperature and is the
    value used for fitting.}
\end{table*}

Fitting to these 8 data points, from \SI{22}{\MHz} to \SI{8}{\GHz}, we
find a good fit to the observed excess with the contribution from FRTs
integrated back to redshift \num{6}.

\begin{figure}
  \centering
  \includegraphics[width=\columnwidth]{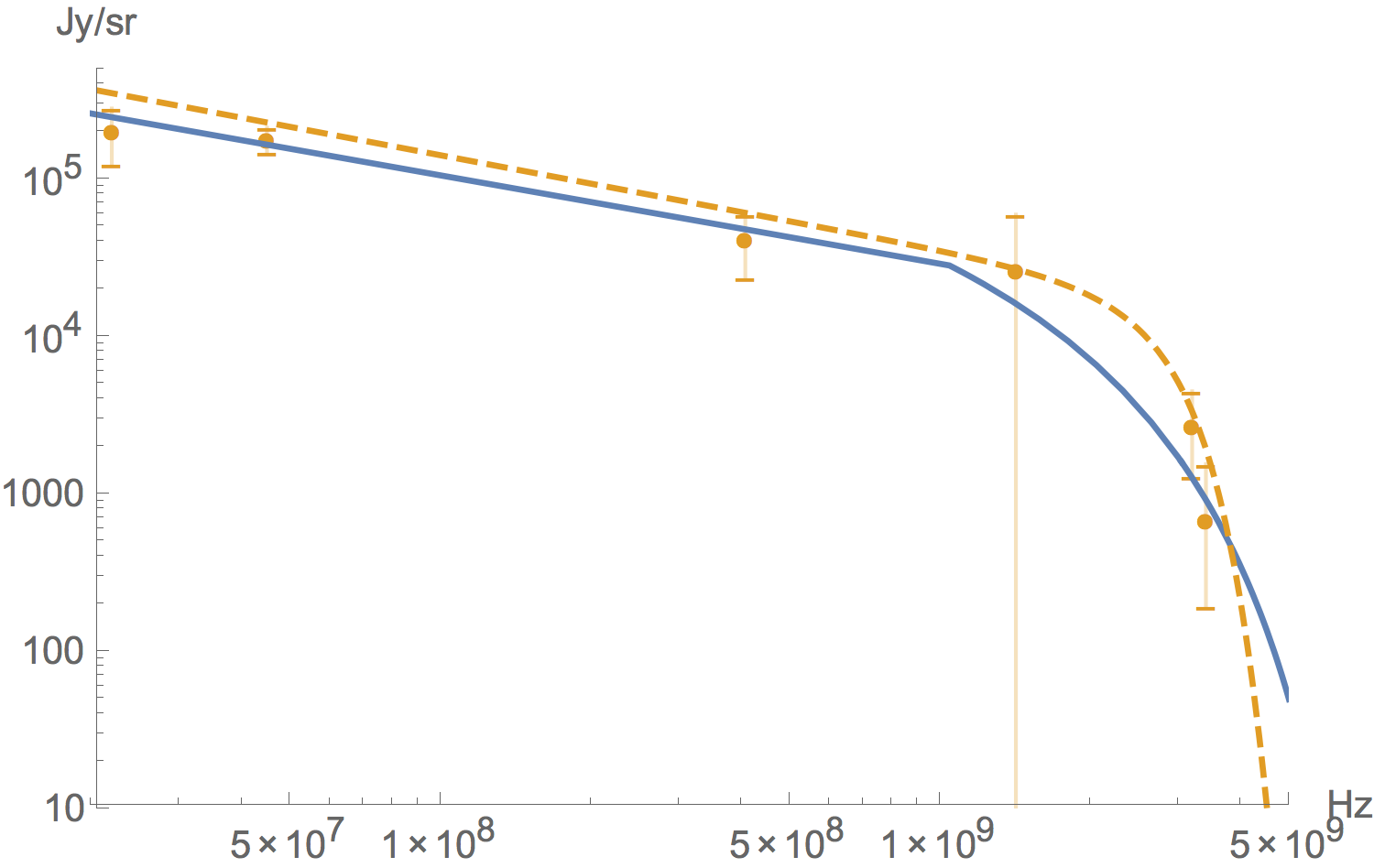}
  \caption{The cosmic radio background due to fast radio transients
    (blue, solid) with cosmological rate based on the comoving volume,
    and the observed radio excess data points (orange) with error bars
    derived from the temperature measurements. The dashed orange line
    is from the continuous best fit function (power law in
    temperature) as reported by ARCADE~2. The excess may continue up
    to about \SI{10}{GHz}, but significance at these high frequencies
    drops and the best fit from the fast transients has no power at
    these high frequencies. Note that at \SI{1.4}{GHz} the error bar
    is consistent with zero, where we've subtracted known radio
    sources as in~\cite{condon}.}
  \label{fig:fit_cv}
\end{figure}

First, we consider the case when we model the transient rate as the
comoving volume. With the parameters for the simple model (the total
luminosity $\cal{L}\times$~rate $R_\text{FRT}$, the spectrum
endpoints, and the spectral index), we find a $\chi^2$ per degree of
freedom of about \num{4.03/4}. The spectral index is \num{0.56} and
the spectrum goes from \SI{10}{\MHz} to \SI{7.3}{\GHz} with a
normalization of \SI{4.6e35}{\J\per\Hz}. Note that the fitted spectrum
cuts off at high frequencies before the final data points at
\num{7.98} and \SI{8.33}{\GHz} reported by ARCADE~2 to have an excess
temperature, but the reported excess of these points are highly
suppressed (of lower significance, and consistent with zero) compared
to the other points. Plots of the fitted model and radio excess are
shown in Figure~\ref{fig:fit_cv}.

\begin{figure}
  \centering
  \includegraphics[width=\columnwidth]{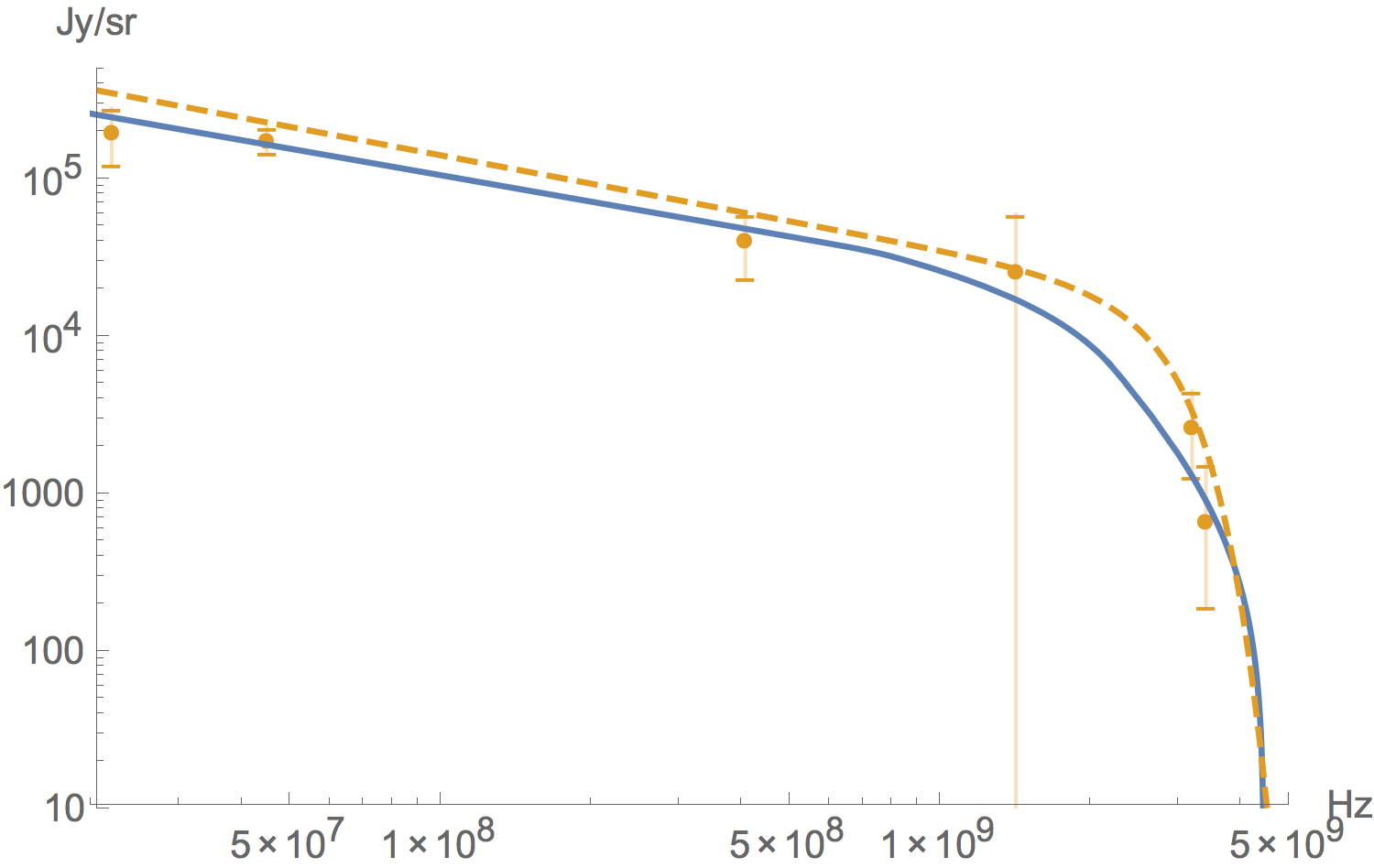}
  \caption{The same as in Figure~\ref{fig:fit_cv} but with the fast
    radio transient rate proportional to the star formation rate.}
  \label{fig:fit_sf}
\end{figure}

For the cosmological fast transient rate modeled after the star
formation rate, the results are similar: The fitted spectrum again has
spectral index \num{0.56}, normalization \SI{4.2e36}{\J\per\Hz}, and
an overall $\chi^2$ per degree of freedom of \num{3.96/4}. The fitted
lower frequency of the spectrum is the same as the previous case, but
the upper endpoint is \SI{4.5}{\GHz}. This model is plotted in
Figure~\ref{fig:fit_sf}. We note that the largest difference between
this background model and the previous one is in the overall
normalization of the spectrum, which needs to be about an order of
magnitude larger in this case.

\section{Discussion}
\label{sec:discussion}

We turn now to interpreting the parameters of the best fit found
above. Compared to the model spectral index of \num{-1.4} for FRBs,
the spectral index of the best fit is very flat in order to match the
observed radio background. And, as we know from the FRB results, we
need an overall factor of about \num{7} orders of magnitude larger to
completely saturate the radio background excess. Alternatively, this
factor needs to be at least \num{5} orders of magnitude larger than
FRBs in order to be at the percent level of the radio excess.

This normalization is the combination of two main factors: the event
rate, and overall energy. One could also subdivide this into further
details, such as possible recurrence rate, luminosity dependence on
redshift (i.e.~a distribution of observed source energies), and so on.

While there have been searches for FRTs in the past (see the recent
review in~\cite{Fender:2015sca}), much of parameter space is
unexplored and unconstrained. Searches are also usually done in narrow
frequency bands, and are not sensitive to general combinations of
parameters (e.g.~minimum peak flux density versus event time). This
makes it particularly difficult to draw general conclusions and
constraints on what FRT parameters are currently allowed, though the
flat fitted spectrum here may already be constrained. Due to the
recent discovery and study of FRBs, this set of parameters (bright
millisecond sources at \SI{1.4}{\GHz}) is better studied. From our
analysis above, it looks rather unlikely that FRBs, or sources very
similar to FRBs, play a role in the ERB\@.

On the other hand, there is still much more to be explored.
Monochromatic sources will be very easy to miss without searching at
exactly the right frequency, for instance. While a single such source
could not explain the entirety of the ERB, perhaps multiple types of
sources play some part. It is not difficult to find parameters for
FRTs that avoid current constraints (which, we stress, are typically
not very general), especially rare very bright sources (perhaps
peaking at unexplored frequencies), or very dim sources. A
comprehensive analysis of the available parameter space, and potential
FRT models, is outside the scope of this current work, but is a
promising future direction. With instruments such as the Square
Kilometer Array aiming to search for radio transients, the coming
years may yield many new discoveries.

\section{Conclusions}
\label{sec:conclusions}

In this work our primary aim has been to extend the analysis of
sources which can contribute to a radio background to include FRTs.
These have not been considered before in relation to the cosmic radio
background. With large swaths of parameter space available, FRTs may
play a role in the ARCADE~2 (and older experiments) excess. To that
end we have used both an estimate based on a differential flux density
and a detailed calculation for transient sources to calculate the
contribution to a radio background.

This analysis has shown that based on what we currently know about
FRBs, they do not contribute at an appreciable level to the radio
excess measurement. One could construct very different source models
which could give a contribution to (or explanation of) the ERB, and
such model building is one possible direction to studying fast
transients and the radio background.

Overall, this work is meant to serve as a first step in a new
direction for transient radio sources and the unknown cosmic radio
background. There is ample opportunity here to propose fast radio
transient models, search and hopefully discover their origin, and
perhaps link them to the observed, yet unexplained, radio background.

\begin{acknowledgments}
  \noindent
  The authors were supported in part by the DOE (DE-SC0011981). TKW
  and TJW thank the Aspen Center for Physics, which is supported by
  NSF grant PHY-1066293, for hospitality while this work was in
  progress.
\end{acknowledgments}

\bibliography{radiobkg_refs}

\begin{thebibliography}{38}%
\makeatletter
\providecommand \@ifxundefined [1]{%
 \@ifx{#1\undefined}
}%
\providecommand \@ifnum [1]{%
 \ifnum #1\expandafter \@firstoftwo
 \else \expandafter \@secondoftwo
 \fi
}%
\providecommand \@ifx [1]{%
 \ifx #1\expandafter \@firstoftwo
 \else \expandafter \@secondoftwo
 \fi
}%
\providecommand \natexlab [1]{#1}%
\providecommand \enquote  [1]{``#1''}%
\providecommand \bibnamefont  [1]{#1}%
\providecommand \bibfnamefont [1]{#1}%
\providecommand \citenamefont [1]{#1}%
\providecommand \href@noop [0]{\@secondoftwo}%
\providecommand \href [0]{\begingroup \@sanitize@url \@href}%
\providecommand \@href[1]{\@@startlink{#1}\@@href}%
\providecommand \@@href[1]{\endgroup#1\@@endlink}%
\providecommand \@sanitize@url [0]{\catcode `\\12\catcode `\$12\catcode
  `\&12\catcode `\#12\catcode `\^12\catcode `\_12\catcode `\%12\relax}%
\providecommand \@@startlink[1]{}%
\providecommand \@@endlink[0]{}%
\providecommand \url  [0]{\begingroup\@sanitize@url \@url }%
\providecommand \@url [1]{\endgroup\@href {#1}{\urlprefix }}%
\providecommand \urlprefix  [0]{URL }%
\providecommand \Eprint [0]{\href }%
\providecommand \doibase [0]{http://dx.doi.org/}%
\providecommand \selectlanguage [0]{\@gobble}%
\providecommand \bibinfo  [0]{\@secondoftwo}%
\providecommand \bibfield  [0]{\@secondoftwo}%
\providecommand \translation [1]{[#1]}%
\providecommand \BibitemOpen [0]{}%
\providecommand \bibitemStop [0]{}%
\providecommand \bibitemNoStop [0]{.\EOS\space}%
\providecommand \EOS [0]{\spacefactor3000\relax}%
\providecommand \BibitemShut  [1]{\csname bibitem#1\endcsname}%
\let\auto@bib@innerbib\@empty
\bibitem [{\citenamefont {Fixsen}\ \emph {et~al.}(2011)\citenamefont {Fixsen},
  \citenamefont {Kogut}, \citenamefont {Levin}, \citenamefont {Limon},
  \citenamefont {Lubin}, \citenamefont {Mirel}, \citenamefont {Seiffert},
  \citenamefont {Singal}, \citenamefont {Wollack}, \citenamefont {Villela},\
  and\ \citenamefont {Wuensche}}]{arcade2}%
  \BibitemOpen
  \bibfield  {author} {\bibinfo {author} {\bibfnamefont {D.~J.}\ \bibnamefont
  {Fixsen}}, \bibinfo {author} {\bibfnamefont {A.}~\bibnamefont {Kogut}},
  \bibinfo {author} {\bibfnamefont {S.}~\bibnamefont {Levin}}, \bibinfo
  {author} {\bibfnamefont {M.}~\bibnamefont {Limon}}, \bibinfo {author}
  {\bibfnamefont {P.}~\bibnamefont {Lubin}}, \bibinfo {author} {\bibfnamefont
  {P.}~\bibnamefont {Mirel}}, \bibinfo {author} {\bibfnamefont
  {M.}~\bibnamefont {Seiffert}}, \bibinfo {author} {\bibfnamefont
  {J.}~\bibnamefont {Singal}}, \bibinfo {author} {\bibfnamefont
  {E.}~\bibnamefont {Wollack}}, \bibinfo {author} {\bibfnamefont
  {T.}~\bibnamefont {Villela}}, \ and\ \bibinfo {author} {\bibfnamefont
  {C.~A.}\ \bibnamefont {Wuensche}},\ }\bibfield  {title} {\enquote {\bibinfo
  {title} {Arcade 2 measurement of the absolute sky brightness at 3-90 ghz},}\
  }\href {http://stacks.iop.org/0004-637X/734/i=1/a=5} {\bibfield  {journal}
  {\bibinfo  {journal} {The Astrophysical Journal}\ }\textbf {\bibinfo {volume}
  {734}},\ \bibinfo {pages} {5} (\bibinfo {year} {2011})}\BibitemShut {NoStop}%
\bibitem [{\citenamefont {Singal}\ \emph {et~al.}(2010)\citenamefont {Singal},
  \citenamefont {Stawarz}, \citenamefont {Lawrence},\ and\ \citenamefont
  {Petrosian}}]{Singal11122010}%
  \BibitemOpen
  \bibfield  {author} {\bibinfo {author} {\bibfnamefont {J.}~\bibnamefont
  {Singal}}, \bibinfo {author} {\bibfnamefont {Ł.}\ \bibnamefont {Stawarz}},
  \bibinfo {author} {\bibfnamefont {A.}~\bibnamefont {Lawrence}}, \ and\
  \bibinfo {author} {\bibfnamefont {V.}~\bibnamefont {Petrosian}},\ }\bibfield
  {title} {\enquote {\bibinfo {title} {Sources of the radio background
  considered},}\ }\href {\doibase 10.1111/j.1365-2966.2010.17382.x} {\bibfield
  {journal} {\bibinfo  {journal} {Monthly Notices of the Royal Astronomical
  Society}\ }\textbf {\bibinfo {volume} {409}},\ \bibinfo {pages} {1172--1182}
  (\bibinfo {year} {2010})}\BibitemShut {NoStop}%
\bibitem [{\citenamefont {Seiffert}\ \emph {et~al.}(2011)\citenamefont
  {Seiffert}, \citenamefont {Fixsen}, \citenamefont {Kogut}, \citenamefont
  {Levin}, \citenamefont {Limon}, \citenamefont {Lubin}, \citenamefont {Mirel},
  \citenamefont {Singal}, \citenamefont {Villela}, \citenamefont {Wollack},\
  and\ \citenamefont {Wuensche}}]{seiffert}%
  \BibitemOpen
  \bibfield  {author} {\bibinfo {author} {\bibfnamefont {M.}~\bibnamefont
  {Seiffert}}, \bibinfo {author} {\bibfnamefont {D.~J.}\ \bibnamefont
  {Fixsen}}, \bibinfo {author} {\bibfnamefont {A.}~\bibnamefont {Kogut}},
  \bibinfo {author} {\bibfnamefont {S.~M.}\ \bibnamefont {Levin}}, \bibinfo
  {author} {\bibfnamefont {M.}~\bibnamefont {Limon}}, \bibinfo {author}
  {\bibfnamefont {P.~M.}\ \bibnamefont {Lubin}}, \bibinfo {author}
  {\bibfnamefont {P.}~\bibnamefont {Mirel}}, \bibinfo {author} {\bibfnamefont
  {J.}~\bibnamefont {Singal}}, \bibinfo {author} {\bibfnamefont
  {T.}~\bibnamefont {Villela}}, \bibinfo {author} {\bibfnamefont
  {E.}~\bibnamefont {Wollack}}, \ and\ \bibinfo {author} {\bibfnamefont
  {C.~A.}\ \bibnamefont {Wuensche}},\ }\bibfield  {title} {\enquote {\bibinfo
  {title} {Interpretation of the arcade 2 absolute sky brightness
  measurement},}\ }\href {http://stacks.iop.org/0004-637X/734/i=1/a=6}
  {\bibfield  {journal} {\bibinfo  {journal} {The Astrophysical Journal}\
  }\textbf {\bibinfo {volume} {734}},\ \bibinfo {pages} {6} (\bibinfo {year}
  {2011})}\BibitemShut {NoStop}%
\bibitem [{\citenamefont {Fornengo}\ \emph {et~al.}(2014)\citenamefont
  {Fornengo}, \citenamefont {Lineros}, \citenamefont {Regis},\ and\
  \citenamefont {Taoso}}]{radiobkgrevist}%
  \BibitemOpen
  \bibfield  {author} {\bibinfo {author} {\bibfnamefont {Nicolao}\ \bibnamefont
  {Fornengo}}, \bibinfo {author} {\bibfnamefont {Roberto~A.}\ \bibnamefont
  {Lineros}}, \bibinfo {author} {\bibfnamefont {Marco}\ \bibnamefont {Regis}},
  \ and\ \bibinfo {author} {\bibfnamefont {Marco}\ \bibnamefont {Taoso}},\
  }\bibfield  {title} {\enquote {\bibinfo {title} {The isotropic radio
  background revisited},}\ }\href
  {http://stacks.iop.org/1475-7516/2014/i=04/a=008} {\bibfield  {journal}
  {\bibinfo  {journal} {Journal of Cosmology and Astroparticle Physics}\
  }\textbf {\bibinfo {volume} {2014}},\ \bibinfo {pages} {008} (\bibinfo {year}
  {2014})}\BibitemShut {NoStop}%
\bibitem [{\citenamefont {Vernstrom}\ \emph {et~al.}(2011)\citenamefont
  {Vernstrom}, \citenamefont {Scott},\ and\ \citenamefont
  {Wall}}]{Vernstrom21082011}%
  \BibitemOpen
  \bibfield  {author} {\bibinfo {author} {\bibfnamefont {T.}~\bibnamefont
  {Vernstrom}}, \bibinfo {author} {\bibfnamefont {Douglas}\ \bibnamefont
  {Scott}}, \ and\ \bibinfo {author} {\bibfnamefont {J.~V.}\ \bibnamefont
  {Wall}},\ }\bibfield  {title} {\enquote {\bibinfo {title} {Contribution to
  the diffuse radio background from extragalactic radio sources},}\ }\href
  {\doibase 10.1111/j.1365-2966.2011.18990.x} {\bibfield  {journal} {\bibinfo
  {journal} {Monthly Notices of the Royal Astronomical Society}\ }\textbf
  {\bibinfo {volume} {415}},\ \bibinfo {pages} {3641--3648} (\bibinfo {year}
  {2011})}\BibitemShut {NoStop}%
\bibitem [{\citenamefont {Holder}(2014)}]{holder}%
  \BibitemOpen
  \bibfield  {author} {\bibinfo {author} {\bibfnamefont {Gilbert~P.}\
  \bibnamefont {Holder}},\ }\bibfield  {title} {\enquote {\bibinfo {title} {The
  unusual smoothness of the extragalactic unresolved radio background},}\
  }\href {http://stacks.iop.org/0004-637X/780/i=1/a=112} {\bibfield  {journal}
  {\bibinfo  {journal} {The Astrophysical Journal}\ }\textbf {\bibinfo {volume}
  {780}},\ \bibinfo {pages} {112} (\bibinfo {year} {2014})}\BibitemShut
  {NoStop}%
\bibitem [{\citenamefont {Fornengo}\ \emph {et~al.}(2011)\citenamefont
  {Fornengo}, \citenamefont {Lineros}, \citenamefont {Regis},\ and\
  \citenamefont {Taoso}}]{arcadebkgmodel1}%
  \BibitemOpen
  \bibfield  {author} {\bibinfo {author} {\bibfnamefont {N.}~\bibnamefont
  {Fornengo}}, \bibinfo {author} {\bibfnamefont {R.}~\bibnamefont {Lineros}},
  \bibinfo {author} {\bibfnamefont {M.}~\bibnamefont {Regis}}, \ and\ \bibinfo
  {author} {\bibfnamefont {M.}~\bibnamefont {Taoso}},\ }\bibfield  {title}
  {\enquote {\bibinfo {title} {{Possibility of a Dark Matter Interpretation for
  the Excess in Isotropic Radio Emission Reported by ARCADE}},}\ }\href
  {\doibase 10.1103/PhysRevLett.107.271302} {\bibfield  {journal} {\bibinfo
  {journal} {Phys.Rev.Lett.}\ }\textbf {\bibinfo {volume} {107}},\ \bibinfo
  {pages} {271302} (\bibinfo {year} {2011})},\ \Eprint
  {http://arxiv.org/abs/1108.0569} {arXiv:1108.0569 [hep-ph]} \BibitemShut
  {NoStop}%
\bibitem [{\citenamefont {Fornengo}\ \emph {et~al.}(2012)\citenamefont
  {Fornengo}, \citenamefont {Lineros}, \citenamefont {Regis},\ and\
  \citenamefont {Taoso}}]{arcadebkgmodel2}%
  \BibitemOpen
  \bibfield  {author} {\bibinfo {author} {\bibfnamefont {N.}~\bibnamefont
  {Fornengo}}, \bibinfo {author} {\bibfnamefont {R.}~\bibnamefont {Lineros}},
  \bibinfo {author} {\bibfnamefont {M.}~\bibnamefont {Regis}}, \ and\ \bibinfo
  {author} {\bibfnamefont {M.}~\bibnamefont {Taoso}},\ }\bibfield  {title}
  {\enquote {\bibinfo {title} {{Cosmological Radio Emission induced by WIMP
  Dark Matter}},}\ }\href {\doibase 10.1088/1475-7516/2012/03/033} {\bibfield
  {journal} {\bibinfo  {journal} {JCAP}\ }\textbf {\bibinfo {volume} {1203}},\
  \bibinfo {pages} {033} (\bibinfo {year} {2012})},\ \Eprint
  {http://arxiv.org/abs/1112.4517} {arXiv:1112.4517 [astro-ph.CO]} \BibitemShut
  {NoStop}%
\bibitem [{\citenamefont {Hooper}\ \emph {et~al.}(2012)\citenamefont {Hooper},
  \citenamefont {Belikov}, \citenamefont {Jeltema}, \citenamefont {Linden},
  \citenamefont {Profumo} \emph {et~al.}}]{arcadebkgmodel3}%
  \BibitemOpen
  \bibfield  {author} {\bibinfo {author} {\bibfnamefont {Dan}\ \bibnamefont
  {Hooper}}, \bibinfo {author} {\bibfnamefont {Alexander~V.}\ \bibnamefont
  {Belikov}}, \bibinfo {author} {\bibfnamefont {Tesla~E.}\ \bibnamefont
  {Jeltema}}, \bibinfo {author} {\bibfnamefont {Tim}\ \bibnamefont {Linden}},
  \bibinfo {author} {\bibfnamefont {Stefano}\ \bibnamefont {Profumo}},  \emph
  {et~al.},\ }\bibfield  {title} {\enquote {\bibinfo {title} {{The Isotropic
  Radio Background and Annihilating Dark Matter}},}\ }\href {\doibase
  10.1103/PhysRevD.86.103003} {\bibfield  {journal} {\bibinfo  {journal}
  {Phys.Rev.}\ }\textbf {\bibinfo {volume} {D86}},\ \bibinfo {pages} {103003}
  (\bibinfo {year} {2012})},\ \Eprint {http://arxiv.org/abs/1203.3547}
  {arXiv:1203.3547 [astro-ph.CO]} \BibitemShut {NoStop}%
\bibitem [{\citenamefont {Yang}\ \emph {et~al.}(2013)\citenamefont {Yang},
  \citenamefont {Yang}, \citenamefont {Huang}, \citenamefont {Chen},
  \citenamefont {Lu} \emph {et~al.}}]{arcadebkgmodel4}%
  \BibitemOpen
  \bibfield  {author} {\bibinfo {author} {\bibfnamefont {Yupeng}\ \bibnamefont
  {Yang}}, \bibinfo {author} {\bibfnamefont {Guilin}\ \bibnamefont {Yang}},
  \bibinfo {author} {\bibfnamefont {Xiaoyuan}\ \bibnamefont {Huang}}, \bibinfo
  {author} {\bibfnamefont {Xuelei}\ \bibnamefont {Chen}}, \bibinfo {author}
  {\bibfnamefont {Tan}\ \bibnamefont {Lu}},  \emph {et~al.},\ }\bibfield
  {title} {\enquote {\bibinfo {title} {{Contribution of ultracompact dark
  matter minihalos to the isotropic radio background}},}\ }\href {\doibase
  10.1103/PhysRevD.87.083519} {\bibfield  {journal} {\bibinfo  {journal}
  {Phys.Rev.}\ }\textbf {\bibinfo {volume} {D87}},\ \bibinfo {pages} {083519}
  (\bibinfo {year} {2013})},\ \Eprint {http://arxiv.org/abs/1206.3750}
  {arXiv:1206.3750 [astro-ph.HE]} \BibitemShut {NoStop}%
\bibitem [{\citenamefont {Lawson}\ and\ \citenamefont
  {Zhitnitsky}(2013)}]{arcadebkgmodel5}%
  \BibitemOpen
  \bibfield  {author} {\bibinfo {author} {\bibfnamefont {Kyle}\ \bibnamefont
  {Lawson}}\ and\ \bibinfo {author} {\bibfnamefont {Ariel~R.}\ \bibnamefont
  {Zhitnitsky}},\ }\bibfield  {title} {\enquote {\bibinfo {title} {{Isotropic
  Radio Background from Quark Nugget Dark Matter}},}\ }\href {\doibase
  10.1016/j.physletb.2013.05.070} {\bibfield  {journal} {\bibinfo  {journal}
  {Phys.Lett.}\ }\textbf {\bibinfo {volume} {B724}},\ \bibinfo {pages} {17--21}
  (\bibinfo {year} {2013})},\ \Eprint {http://arxiv.org/abs/1210.2400}
  {arXiv:1210.2400 [astro-ph.CO]} \BibitemShut {NoStop}%
\bibitem [{\citenamefont {Cline}\ and\ \citenamefont
  {Vincent}(2013)}]{arcadebkgmodel6}%
  \BibitemOpen
  \bibfield  {author} {\bibinfo {author} {\bibfnamefont {James~M.}\
  \bibnamefont {Cline}}\ and\ \bibinfo {author} {\bibfnamefont {Aaron~C.}\
  \bibnamefont {Vincent}},\ }\bibfield  {title} {\enquote {\bibinfo {title}
  {{Cosmological origin of anomalous radio background}},}\ }\href {\doibase
  10.1088/1475-7516/2013/02/011} {\bibfield  {journal} {\bibinfo  {journal}
  {JCAP}\ }\textbf {\bibinfo {volume} {1302}},\ \bibinfo {pages} {011}
  (\bibinfo {year} {2013})},\ \Eprint {http://arxiv.org/abs/1210.2717}
  {arXiv:1210.2717 [astro-ph.CO]} \BibitemShut {NoStop}%
\bibitem [{\citenamefont {Lorimer}\ \emph {et~al.}(2007)\citenamefont
  {Lorimer}, \citenamefont {Bailes}, \citenamefont {McLaughlin}, \citenamefont
  {Narkevic},\ and\ \citenamefont {Crawford}}]{parkes1}%
  \BibitemOpen
  \bibfield  {author} {\bibinfo {author} {\bibfnamefont {D.~R.}\ \bibnamefont
  {Lorimer}}, \bibinfo {author} {\bibfnamefont {M.}~\bibnamefont {Bailes}},
  \bibinfo {author} {\bibfnamefont {M.~A.}\ \bibnamefont {McLaughlin}},
  \bibinfo {author} {\bibfnamefont {D.~J.}\ \bibnamefont {Narkevic}}, \ and\
  \bibinfo {author} {\bibfnamefont {F.}~\bibnamefont {Crawford}},\ }\bibfield
  {title} {\enquote {\bibinfo {title} {A bright millisecond radio burst of
  extragalactic origin},}\ }\href {\doibase 10.1126/science.1147532} {\bibfield
   {journal} {\bibinfo  {journal} {Science}\ }\textbf {\bibinfo {volume}
  {318}},\ \bibinfo {pages} {777--780} (\bibinfo {year} {2007})}\BibitemShut
  {NoStop}%
\bibitem [{\citenamefont {Keane}\ \emph {et~al.}(2012)\citenamefont {Keane},
  \citenamefont {Stappers}, \citenamefont {Kramer},\ and\ \citenamefont
  {Lyne}}]{parkes2}%
  \BibitemOpen
  \bibfield  {author} {\bibinfo {author} {\bibfnamefont {E.~F.}\ \bibnamefont
  {Keane}}, \bibinfo {author} {\bibfnamefont {B.~W.}\ \bibnamefont {Stappers}},
  \bibinfo {author} {\bibfnamefont {M.}~\bibnamefont {Kramer}}, \ and\ \bibinfo
  {author} {\bibfnamefont {A.~G.}\ \bibnamefont {Lyne}},\ }\bibfield  {title}
  {\enquote {\bibinfo {title} {On the origin of a highly dispersed coherent
  radio burst},}\ }\href {\doibase 10.1111/j.1745-3933.2012.01306.x} {\bibfield
   {journal} {\bibinfo  {journal} {Monthly Notices of the Royal Astronomical
  Society: Letters}\ }\textbf {\bibinfo {volume} {425}},\ \bibinfo {pages}
  {L71--L75} (\bibinfo {year} {2012})}\BibitemShut {NoStop}%
\bibitem [{\citenamefont {Thornton}\ \emph {et~al.}(2013)\citenamefont
  {Thornton}, \citenamefont {Stappers}, \citenamefont {Bailes}, \citenamefont
  {Barsdell}, \citenamefont {Bates} \emph {et~al.}}]{parkes3}%
  \BibitemOpen
  \bibfield  {author} {\bibinfo {author} {\bibfnamefont {D.}~\bibnamefont
  {Thornton}}, \bibinfo {author} {\bibfnamefont {B.}~\bibnamefont {Stappers}},
  \bibinfo {author} {\bibfnamefont {M.}~\bibnamefont {Bailes}}, \bibinfo
  {author} {\bibfnamefont {B.R.}\ \bibnamefont {Barsdell}}, \bibinfo {author}
  {\bibfnamefont {S.D.}\ \bibnamefont {Bates}},  \emph {et~al.},\ }\bibfield
  {title} {\enquote {\bibinfo {title} {{A Population of Fast Radio Bursts at
  Cosmological Distances}},}\ }\href {\doibase 10.1126/science.1236789}
  {\bibfield  {journal} {\bibinfo  {journal} {Science}\ }\textbf {\bibinfo
  {volume} {341}},\ \bibinfo {pages} {53--56} (\bibinfo {year} {2013})},\
  \Eprint {http://arxiv.org/abs/1307.1628} {arXiv:1307.1628 [astro-ph.HE]}
  \BibitemShut {NoStop}%
\bibitem [{\citenamefont {Spitler}\ \emph {et~al.}(2014)\citenamefont
  {Spitler}, \citenamefont {Cordes}, \citenamefont {Hessels}, \citenamefont
  {Lorimer}, \citenamefont {McLaughlin} \emph {et~al.}}]{arecibo}%
  \BibitemOpen
  \bibfield  {author} {\bibinfo {author} {\bibfnamefont {L.G.}\ \bibnamefont
  {Spitler}}, \bibinfo {author} {\bibfnamefont {J.M.}\ \bibnamefont {Cordes}},
  \bibinfo {author} {\bibfnamefont {J.W.T.}\ \bibnamefont {Hessels}}, \bibinfo
  {author} {\bibfnamefont {D.R.}\ \bibnamefont {Lorimer}}, \bibinfo {author}
  {\bibfnamefont {M.A.}\ \bibnamefont {McLaughlin}},  \emph {et~al.},\
  }\bibfield  {title} {\enquote {\bibinfo {title} {{Fast Radio Burst Discovered
  in the Arecibo Pulsar ALFA Survey}},}\ }\href {\doibase
  10.1088/0004-637X/790/2/101} {\bibfield  {journal} {\bibinfo  {journal}
  {Astrophys.J.}\ }\textbf {\bibinfo {volume} {790}},\ \bibinfo {pages} {101}
  (\bibinfo {year} {2014})},\ \Eprint {http://arxiv.org/abs/1404.2934}
  {arXiv:1404.2934 [astro-ph.HE]} \BibitemShut {NoStop}%
\bibitem [{\citenamefont {Thornton}(2013)}]{thorntonthesis}%
  \BibitemOpen
  \bibfield  {author} {\bibinfo {author} {\bibfnamefont {D.~G.~P.}\
  \bibnamefont {Thornton}},\ }\emph {\bibinfo {title} {The High Time Resolution
  Radio Sky}},\ \href@noop {} {Ph.D. thesis},\ \bibinfo  {school} {The
  University of Manchester} (\bibinfo {year} {2013})\BibitemShut {NoStop}%
\bibitem [{\citenamefont {Burke-Spolaor}\ and\ \citenamefont
  {Bannister}(2014)}]{Burke-Spolaor:2014rqa}%
  \BibitemOpen
  \bibfield  {author} {\bibinfo {author} {\bibfnamefont {Sarah}\ \bibnamefont
  {Burke-Spolaor}}\ and\ \bibinfo {author} {\bibfnamefont {Keith~W.}\
  \bibnamefont {Bannister}},\ }\bibfield  {title} {\enquote {\bibinfo {title}
  {{The Galactic Position Dependence of Fast Radio Bursts and the Discovery of
  FRB011025}},}\ }\href {\doibase 10.1088/0004-637X/792/1/19} {\bibfield
  {journal} {\bibinfo  {journal} {Astrophys. J.}\ }\textbf {\bibinfo {volume}
  {792}},\ \bibinfo {pages} {19} (\bibinfo {year} {2014})},\ \Eprint
  {http://arxiv.org/abs/1407.0400} {arXiv:1407.0400 [astro-ph.HE]} \BibitemShut
  {NoStop}%
\bibitem [{\citenamefont {Petroff}\ \emph {et~al.}(2015)\citenamefont
  {Petroff}, \citenamefont {Bailes}, \citenamefont {Barr}, \citenamefont
  {Barsdell}, \citenamefont {Bhat} \emph {et~al.}}]{Petroff:2014taa}%
  \BibitemOpen
  \bibfield  {author} {\bibinfo {author} {\bibfnamefont {E.}~\bibnamefont
  {Petroff}}, \bibinfo {author} {\bibfnamefont {M.}~\bibnamefont {Bailes}},
  \bibinfo {author} {\bibfnamefont {E.~D.}\ \bibnamefont {Barr}}, \bibinfo
  {author} {\bibfnamefont {B.~R.}\ \bibnamefont {Barsdell}}, \bibinfo {author}
  {\bibfnamefont {N.~D.~R.}\ \bibnamefont {Bhat}},  \emph {et~al.},\ }\bibfield
   {title} {\enquote {\bibinfo {title} {{A real-time fast radio burst:
  polarization detection and multiwavelength follow-up}},}\ }\href {\doibase
  10.1093/mnras/stu2419} {\bibfield  {journal} {\bibinfo  {journal} {Mon. Not.
  Roy. Astron. Soc.}\ }\textbf {\bibinfo {volume} {447}},\ \bibinfo {pages}
  {246--255} (\bibinfo {year} {2015})},\ \Eprint
  {http://arxiv.org/abs/1412.0342} {arXiv:1412.0342 [astro-ph.HE]} \BibitemShut
  {NoStop}%
\bibitem [{\citenamefont {Ravi}\ \emph {et~al.}(2015)\citenamefont {Ravi},
  \citenamefont {Shannon},\ and\ \citenamefont {Jameson}}]{Ravi:2014mma}%
  \BibitemOpen
  \bibfield  {author} {\bibinfo {author} {\bibfnamefont {V.}~\bibnamefont
  {Ravi}}, \bibinfo {author} {\bibfnamefont {R.~M.}\ \bibnamefont {Shannon}}, \
  and\ \bibinfo {author} {\bibfnamefont {A.}~\bibnamefont {Jameson}},\
  }\bibfield  {title} {\enquote {\bibinfo {title} {{A fast radio burst in the
  direction of the Carina dwarf spheroidal galaxy}},}\ }\href {\doibase
  10.1088/2041-8205/799/1/L5} {\bibfield  {journal} {\bibinfo  {journal}
  {Astrophys. J.}\ }\textbf {\bibinfo {volume} {799}},\ \bibinfo {pages} {L5}
  (\bibinfo {year} {2015})},\ \Eprint {http://arxiv.org/abs/1412.1599}
  {arXiv:1412.1599 [astro-ph.HE]} \BibitemShut {NoStop}%
\bibitem [{\citenamefont {Kulkarni}\ \emph {et~al.}(2014)\citenamefont
  {Kulkarni}, \citenamefont {Ofek}, \citenamefont {Neill}, \citenamefont
  {Zheng},\ and\ \citenamefont {Juric}}]{Kulkarni:2014vea}%
  \BibitemOpen
  \bibfield  {author} {\bibinfo {author} {\bibfnamefont {S.~R.}\ \bibnamefont
  {Kulkarni}}, \bibinfo {author} {\bibfnamefont {E.~O.}\ \bibnamefont {Ofek}},
  \bibinfo {author} {\bibfnamefont {J.~D.}\ \bibnamefont {Neill}}, \bibinfo
  {author} {\bibfnamefont {Z.}~\bibnamefont {Zheng}}, \ and\ \bibinfo {author}
  {\bibfnamefont {M.}~\bibnamefont {Juric}},\ }\bibfield  {title} {\enquote
  {\bibinfo {title} {{Giant Sparks at Cosmological Distances?}}}\ }\href
  {\doibase 10.1088/0004-637X/797/1/70} {\bibfield  {journal} {\bibinfo
  {journal} {Astrophys. J.}\ }\textbf {\bibinfo {volume} {797}},\ \bibinfo
  {pages} {70} (\bibinfo {year} {2014})},\ \Eprint
  {http://arxiv.org/abs/1402.4766} {arXiv:1402.4766 [astro-ph.HE]} \BibitemShut
  {NoStop}%
\bibitem [{\citenamefont {{Maoz}}\ \emph {et~al.}(2015)\citenamefont {{Maoz}},
  \citenamefont {{Loeb}}, \citenamefont {{Shvartzvald}}, \citenamefont
  {{Sitek}}, \citenamefont {{Engel}} \emph {et~al.}}]{loebgalactic}%
  \BibitemOpen
  \bibfield  {author} {\bibinfo {author} {\bibfnamefont {D.}~\bibnamefont
  {{Maoz}}}, \bibinfo {author} {\bibfnamefont {A.}~\bibnamefont {{Loeb}}},
  \bibinfo {author} {\bibfnamefont {Y.}~\bibnamefont {{Shvartzvald}}}, \bibinfo
  {author} {\bibfnamefont {M.}~\bibnamefont {{Sitek}}}, \bibinfo {author}
  {\bibfnamefont {M.}~\bibnamefont {{Engel}}},  \emph {et~al.},\ }\bibfield
  {title} {\enquote {\bibinfo {title} {{Fast radio bursts: the observational
  case for a Galactic origin}},}\ }\href@noop {} {\  (\bibinfo {year}
  {2015})},\ \Eprint {http://arxiv.org/abs/1507.01002} {arXiv:1507.01002
  [astro-ph.SR]} \BibitemShut {NoStop}%
\bibitem [{\citenamefont {Totani}(2013)}]{neutron1}%
  \BibitemOpen
  \bibfield  {author} {\bibinfo {author} {\bibfnamefont {Tomonori}\
  \bibnamefont {Totani}},\ }\bibfield  {title} {\enquote {\bibinfo {title}
  {Cosmological fast radio bursts from binary neutron star mergers},}\ }\href
  {\doibase 10.1093/pasj/65.5.L12} {\bibfield  {journal} {\bibinfo  {journal}
  {Publications of the Astronomical Society of Japan}\ }\textbf {\bibinfo
  {volume} {65}} (\bibinfo {year} {2013}),\ 10.1093/pasj/65.5.L12}\BibitemShut
  {NoStop}%
\bibitem [{\citenamefont {Falcke}\ and\ \citenamefont
  {Rezzolla}(2014)}]{neutron2}%
  \BibitemOpen
  \bibfield  {author} {\bibinfo {author} {\bibfnamefont {Heino}\ \bibnamefont
  {Falcke}}\ and\ \bibinfo {author} {\bibfnamefont {Luciano}\ \bibnamefont
  {Rezzolla}},\ }\bibfield  {title} {\enquote {\bibinfo {title} {{Fast radio
  bursts: the last sign of supramassive neutron stars}},}\ }\href {\doibase
  10.1051/0004-6361/201321996} {\bibfield  {journal} {\bibinfo  {journal}
  {Astron. Astrophys.}\ }\textbf {\bibinfo {volume} {562}},\ \bibinfo {pages}
  {A137} (\bibinfo {year} {2014})},\ \Eprint {http://arxiv.org/abs/1307.1409}
  {arXiv:1307.1409 [astro-ph.HE]} \BibitemShut {NoStop}%
\bibitem [{\citenamefont {Iwazaki}(2015)}]{axionfrb1}%
  \BibitemOpen
  \bibfield  {author} {\bibinfo {author} {\bibfnamefont {A.}~\bibnamefont
  {Iwazaki}},\ }\bibfield  {title} {\enquote {\bibinfo {title} {{Axion stars
  and fast radio bursts}},}\ }\href {\doibase 10.1103/PhysRevD.91.023008}
  {\bibfield  {journal} {\bibinfo  {journal} {Phys. Rev.}\ }\textbf {\bibinfo
  {volume} {D91}},\ \bibinfo {pages} {023008} (\bibinfo {year} {2015})},\
  \Eprint {http://arxiv.org/abs/1410.4323} {arXiv:1410.4323 [hep-ph]}
  \BibitemShut {NoStop}%
\bibitem [{\citenamefont {Tkachev}(2015)}]{axionfrb2}%
  \BibitemOpen
  \bibfield  {author} {\bibinfo {author} {\bibfnamefont {I.~I.}\ \bibnamefont
  {Tkachev}},\ }\bibfield  {title} {\enquote {\bibinfo {title} {{Fast Radio
  Bursts and Axion Miniclusters}},}\ }\href {\doibase
  10.1134/S0021364015010154} {\bibfield  {journal} {\bibinfo  {journal} {JETP
  Lett.}\ }\textbf {\bibinfo {volume} {101}},\ \bibinfo {pages} {1--6}
  (\bibinfo {year} {2015})},\ \Eprint {http://arxiv.org/abs/1411.3900}
  {arXiv:1411.3900 [astro-ph.HE]} \BibitemShut {NoStop}%
\bibitem [{\citenamefont {Lipunov}\ and\ \citenamefont
  {Pruzhinskaya}(2014)}]{Lipunov:2013axa}%
  \BibitemOpen
  \bibfield  {author} {\bibinfo {author} {\bibfnamefont {V.~M.}\ \bibnamefont
  {Lipunov}}\ and\ \bibinfo {author} {\bibfnamefont {M.~V.}\ \bibnamefont
  {Pruzhinskaya}},\ }\bibfield  {title} {\enquote {\bibinfo {title} {{Scenario
  Machine: fast radio bursts, short gamma-ray burst, dark energy and Laser
  Interferometer Gravitational-wave Observatory silence}},}\ }\href {\doibase
  10.1093/mnras/stu313} {\bibfield  {journal} {\bibinfo  {journal} {Mon. Not.
  Roy. Astron. Soc.}\ }\textbf {\bibinfo {volume} {440}},\ \bibinfo {pages}
  {1193--1199} (\bibinfo {year} {2014})},\ \Eprint
  {http://arxiv.org/abs/1312.3143} {arXiv:1312.3143 [astro-ph.HE]} \BibitemShut
  {NoStop}%
\bibitem [{\citenamefont {Popov}\ and\ \citenamefont
  {Postnov}(2013)}]{Popov:2013bia}%
  \BibitemOpen
  \bibfield  {author} {\bibinfo {author} {\bibfnamefont {S.~B.}\ \bibnamefont
  {Popov}}\ and\ \bibinfo {author} {\bibfnamefont {K.~A.}\ \bibnamefont
  {Postnov}},\ }\bibfield  {title} {\enquote {\bibinfo {title} {{Millisecond
  extragalactic radio bursts as magnetar flares}},}\ }\href@noop {} {\
  (\bibinfo {year} {2013})},\ \Eprint {http://arxiv.org/abs/1307.4924}
  {arXiv:1307.4924 [astro-ph.HE]} \BibitemShut {NoStop}%
\bibitem [{\citenamefont {Kashiyama}\ \emph {et~al.}(2013)\citenamefont
  {Kashiyama}, \citenamefont {Ioka},\ and\ \citenamefont
  {Mészáros}}]{Kashiyama:2013gza}%
  \BibitemOpen
  \bibfield  {author} {\bibinfo {author} {\bibfnamefont {Kazumi}\ \bibnamefont
  {Kashiyama}}, \bibinfo {author} {\bibfnamefont {Kunihito}\ \bibnamefont
  {Ioka}}, \ and\ \bibinfo {author} {\bibfnamefont {Peter}\ \bibnamefont
  {Mészáros}},\ }\bibfield  {title} {\enquote {\bibinfo {title}
  {{Cosmological Fast Radio Bursts from Binary White Dwarf Mergers}},}\ }\href
  {\doibase 10.1088/2041-8205/776/2/L39} {\bibfield  {journal} {\bibinfo
  {journal} {Astrophys. J.}\ }\textbf {\bibinfo {volume} {776}},\ \bibinfo
  {pages} {L39} (\bibinfo {year} {2013})},\ \Eprint
  {http://arxiv.org/abs/1307.7708} {arXiv:1307.7708 [astro-ph.HE]} \BibitemShut
  {NoStop}%
\bibitem [{\citenamefont {Connor}\ \emph {et~al.}(2015)\citenamefont {Connor},
  \citenamefont {Sievers},\ and\ \citenamefont {Pen}}]{Connor:2015era}%
  \BibitemOpen
  \bibfield  {author} {\bibinfo {author} {\bibfnamefont {Liam}\ \bibnamefont
  {Connor}}, \bibinfo {author} {\bibfnamefont {Jonathan}\ \bibnamefont
  {Sievers}}, \ and\ \bibinfo {author} {\bibfnamefont {Ue-Li}\ \bibnamefont
  {Pen}},\ }\bibfield  {title} {\enquote {\bibinfo {title} {{Non-Cosmological
  FRB's from Young Supernova Remnant Pulsars}},}\ }\href@noop {} {\  (\bibinfo
  {year} {2015})},\ \Eprint {http://arxiv.org/abs/1505.05535} {arXiv:1505.05535
  [astro-ph.HE]} \BibitemShut {NoStop}%
\bibitem [{\citenamefont {Katz}(2015)}]{Katz:2015mpa}%
  \BibitemOpen
  \bibfield  {author} {\bibinfo {author} {\bibfnamefont {J.~I.}\ \bibnamefont
  {Katz}},\ }\bibfield  {title} {\enquote {\bibinfo {title} {{Inferences from
  the Distributions of Fast Radio Burst Pulse Widths, Dispersion Measures and
  Fluences}},}\ }\href@noop {} {\  (\bibinfo {year} {2015})},\ \Eprint
  {http://arxiv.org/abs/1505.06220} {arXiv:1505.06220 [astro-ph.HE]}
  \BibitemShut {NoStop}%
\bibitem [{\citenamefont {Rane}\ \emph {et~al.}(2015)\citenamefont {Rane},
  \citenamefont {Lorimer}, \citenamefont {Bates}, \citenamefont {McMann},
  \citenamefont {McLaughlin},\ and\ \citenamefont {Rajwade}}]{Rane:2015sxa}%
  \BibitemOpen
  \bibfield  {author} {\bibinfo {author} {\bibfnamefont {A.}~\bibnamefont
  {Rane}}, \bibinfo {author} {\bibfnamefont {D.~R.}\ \bibnamefont {Lorimer}},
  \bibinfo {author} {\bibfnamefont {S.~D.}\ \bibnamefont {Bates}}, \bibinfo
  {author} {\bibfnamefont {N.}~\bibnamefont {McMann}}, \bibinfo {author}
  {\bibfnamefont {M.~A.}\ \bibnamefont {McLaughlin}}, \ and\ \bibinfo {author}
  {\bibfnamefont {K.}~\bibnamefont {Rajwade}},\ }\bibfield  {title} {\enquote
  {\bibinfo {title} {{A search for rotating radio transients and fast radio
  bursts in the Parkes high-latitude pulsar survey}},}\ }\href@noop {} {\
  (\bibinfo {year} {2015})},\ \Eprint {http://arxiv.org/abs/1505.00834}
  {arXiv:1505.00834 [astro-ph.HE]} \BibitemShut {NoStop}%
\bibitem [{\citenamefont {Fender}\ \emph {et~al.}(2015)\citenamefont {Fender},
  \citenamefont {Stewart}, \citenamefont {Macquart}, \citenamefont
  {Donnarumma}, \citenamefont {Murphy} \emph {et~al.}}]{Fender:2015sca}%
  \BibitemOpen
  \bibfield  {author} {\bibinfo {author} {\bibfnamefont {Rob}\ \bibnamefont
  {Fender}}, \bibinfo {author} {\bibfnamefont {Adam}\ \bibnamefont {Stewart}},
  \bibinfo {author} {\bibfnamefont {Jean-Pierre}\ \bibnamefont {Macquart}},
  \bibinfo {author} {\bibfnamefont {Immacolata}\ \bibnamefont {Donnarumma}},
  \bibinfo {author} {\bibfnamefont {Tara}\ \bibnamefont {Murphy}},  \emph
  {et~al.},\ }\bibfield  {title} {\enquote {\bibinfo {title} {{Transient
  Astrophysics with the Square Kilometre Array}},}\ }\href@noop {} {\
  (\bibinfo {year} {2015})},\ \Eprint {http://arxiv.org/abs/1507.00729}
  {arXiv:1507.00729 [astro-ph.HE]} \BibitemShut {NoStop}%
\bibitem [{\citenamefont {Lunardini}(2010)}]{Lunardini:2010ab}%
  \BibitemOpen
  \bibfield  {author} {\bibinfo {author} {\bibfnamefont {Cecilia}\ \bibnamefont
  {Lunardini}},\ }\bibfield  {title} {\enquote {\bibinfo {title} {{Diffuse
  supernova neutrinos at underground laboratories}},}\ }\href@noop {} {\
  (\bibinfo {year} {2010})},\ \Eprint {http://arxiv.org/abs/1007.3252}
  {arXiv:1007.3252 [astro-ph.CO]} \BibitemShut {NoStop}%
\bibitem [{\citenamefont {Lorimer}\ \emph {et~al.}(2013)\citenamefont
  {Lorimer}, \citenamefont {Karastergiou}, \citenamefont {McLaughlin},\ and\
  \citenamefont {Johnston}}]{Lorimer21112013}%
  \BibitemOpen
  \bibfield  {author} {\bibinfo {author} {\bibfnamefont {D.~R.}\ \bibnamefont
  {Lorimer}}, \bibinfo {author} {\bibfnamefont {A.}~\bibnamefont
  {Karastergiou}}, \bibinfo {author} {\bibfnamefont {M.~A.}\ \bibnamefont
  {McLaughlin}}, \ and\ \bibinfo {author} {\bibfnamefont {S.}~\bibnamefont
  {Johnston}},\ }\bibfield  {title} {\enquote {\bibinfo {title} {On the
  detectability of extragalactic fast radio transients},}\ }\href {\doibase
  10.1093/mnrasl/slt098} {\bibfield  {journal} {\bibinfo  {journal} {Monthly
  Notices of the Royal Astronomical Society: Letters}\ }\textbf {\bibinfo
  {volume} {436}},\ \bibinfo {pages} {L5--L9} (\bibinfo {year}
  {2013})}\BibitemShut {NoStop}%
\bibitem [{\citenamefont {Hopkins}\ and\ \citenamefont
  {Beacom}(2006)}]{Hopkins:2006bw}%
  \BibitemOpen
  \bibfield  {author} {\bibinfo {author} {\bibfnamefont {Andrew~M.}\
  \bibnamefont {Hopkins}}\ and\ \bibinfo {author} {\bibfnamefont {John~F.}\
  \bibnamefont {Beacom}},\ }\bibfield  {title} {\enquote {\bibinfo {title} {{On
  the normalisation of the cosmic star formation history}},}\ }\href {\doibase
  10.1086/506610} {\bibfield  {journal} {\bibinfo  {journal} {Astrophys. J.}\
  }\textbf {\bibinfo {volume} {651}},\ \bibinfo {pages} {142--154} (\bibinfo
  {year} {2006})},\ \Eprint {http://arxiv.org/abs/astro-ph/0601463}
  {arXiv:astro-ph/0601463 [astro-ph]} \BibitemShut {NoStop}%
\bibitem [{\citenamefont {Gervasi}\ \emph {et~al.}(2008)\citenamefont
  {Gervasi}, \citenamefont {Tartari}, \citenamefont {Zannoni}, \citenamefont
  {Boella},\ and\ \citenamefont {Sironi}}]{gervasi}%
  \BibitemOpen
  \bibfield  {author} {\bibinfo {author} {\bibfnamefont {M.}~\bibnamefont
  {Gervasi}}, \bibinfo {author} {\bibfnamefont {A.}~\bibnamefont {Tartari}},
  \bibinfo {author} {\bibfnamefont {M.}~\bibnamefont {Zannoni}}, \bibinfo
  {author} {\bibfnamefont {G.}~\bibnamefont {Boella}}, \ and\ \bibinfo {author}
  {\bibfnamefont {G.}~\bibnamefont {Sironi}},\ }\bibfield  {title} {\enquote
  {\bibinfo {title} {The contribution of the unresolved extragalactic radio
  sources to the brightness temperature of the sky},}\ }\href
  {http://stacks.iop.org/0004-637X/682/i=1/a=223} {\bibfield  {journal}
  {\bibinfo  {journal} {The Astrophysical Journal}\ }\textbf {\bibinfo {volume}
  {682}},\ \bibinfo {pages} {223} (\bibinfo {year} {2008})}\BibitemShut
  {NoStop}%
\bibitem [{\citenamefont {Condon}\ \emph {et~al.}(2012)\citenamefont {Condon},
  \citenamefont {Cotton}, \citenamefont {Fomalont}, \citenamefont {Kellermann},
  \citenamefont {Miller}, \citenamefont {Perley}, \citenamefont {Scott},
  \citenamefont {Vernstrom},\ and\ \citenamefont {Wall}}]{condon}%
  \BibitemOpen
  \bibfield  {author} {\bibinfo {author} {\bibfnamefont {J.~J.}\ \bibnamefont
  {Condon}}, \bibinfo {author} {\bibfnamefont {W.~D.}\ \bibnamefont {Cotton}},
  \bibinfo {author} {\bibfnamefont {E.~B.}\ \bibnamefont {Fomalont}}, \bibinfo
  {author} {\bibfnamefont {K.~I.}\ \bibnamefont {Kellermann}}, \bibinfo
  {author} {\bibfnamefont {N.}~\bibnamefont {Miller}}, \bibinfo {author}
  {\bibfnamefont {R.~A.}\ \bibnamefont {Perley}}, \bibinfo {author}
  {\bibfnamefont {D.}~\bibnamefont {Scott}}, \bibinfo {author} {\bibfnamefont
  {T.}~\bibnamefont {Vernstrom}}, \ and\ \bibinfo {author} {\bibfnamefont
  {J.~V.}\ \bibnamefont {Wall}},\ }\bibfield  {title} {\enquote {\bibinfo
  {title} {Resolving the radio source background: Deeper understanding through
  confusion},}\ }\href {http://stacks.iop.org/0004-637X/758/i=1/a=23}
  {\bibfield  {journal} {\bibinfo  {journal} {The Astrophysical Journal}\
  }\textbf {\bibinfo {volume} {758}},\ \bibinfo {pages} {23} (\bibinfo {year}
  {2012})}\BibitemShut {NoStop}%
\end{thebibliography}%

\end{document}